\documentclass [10pt,A4paper,conference]{IEEEtran}
\usepackage{flushend}
\usepackage{graphicx}
\begin{document}
\title{On the Spread of Random Interleavers}
\author{\authorblockN{Arya Mazumdar, Adrish Banerjee and A K Chaturvedi}
\authorblockA{Department of Electrical Engineering\\
Indian Institute of Technology\\
Kanpur, India \\
Email: \{arya, adrish, akc\}@iitk.ac.in}
 }
 
\maketitle

\begin{abstract}
 For a given blocklength we determine the  number of interleavers which have spread equal to two. Using this, we find out the  probability that a randomly chosen interleaver has spread two. We show that as blocklength increases, this probability increases but very quickly converges to the value  $1-e^{-2} \approx 0.8647$. Subsequently, we determine a lower bound on the probability of an interleaver having  spread at least $s$. We  show that this lower bound  converges to the
 value $e^{-2(s-2)^{2}}$, as the blocklength increases.
\end{abstract}

\quad

\section{Introduction}

\quad

Interleavers play an important role in digital communications over fading channels. They are also a key component of turbo codes \cite{beurro}\cite{benedetto}\cite{benedetto1}. 
A key characteristic of an interleaver is its spread. The notion of spread  was initially introduced in \cite{divsalar} to take into account the short cycles \cite{costello}\cite{hok} which occur  when
two bits that are
initially close to each other, remain close after interleaving. 

Spread has been
redefined in \cite{Spr} in the following way. Say, $\pi$ is a permutation on the set \{0,1,2....,N-1\}. The spread $s$ of an interleaver is defined as,
\begin{eqnarray}
s\quad \stackrel{\bigtriangleup}{=} \quad {\min_{i,j}}
\left(\left|i-j\right|_{N}\  +\left|\pi (i)-(\pi (j)\right|_{N}\right)\  \nonumber\\
\quad 0\leq i,j\leq N-1,i\ne j
\end{eqnarray}
where
\begin{equation}
|a-b|_{N}=\min(|a-b|,N-|a-b|).
\end{equation}

It is known that the maximum  possible spread for an  interleaver with blocklength $N$ is
$\lfloor \sqrt{2N}\rfloor$  \cite{divsalar}\cite{Spr}\cite{Sphere}. 
This paper seeks to address the question that if we randomly pick up an interleaver from the set of $N!$  possible interleavers, what is the probability that the spread of the interleaver  is $s$, where $2 \leq s\leq \lfloor \sqrt{2N}\rfloor$. We determine this probability for $s=2$ while for other values of $s$ we determine a lower bound on the probability that the spread is at least $s$. We further determine this probability for the limiting case when the blocklength approaches infinity. Using these results it has been possible to determine a lower bound on the number of interleavers having a spread at least $s$ for a  given blocklength.


 The paper is organized as follows. In section II, a  combinatorial analysis is done with the help of \cite{Abra} to find out the number of  interleavers with spread two.
The probability of a random interleaver having a spread two is also derived.
In sections III and IV, a probabilistic analysis has been done to determine a lower bound on the probability that a randomly chosen interleaver will have  spread at least s. We conclude the paper in section V.

\quad
\vspace{-0.3cm}

\section{Interleavers With Spread Two}

\quad


 
An interleaver will have a spread of 2 if at least one pair of consecutive positions remains neighbors after interleaving. We will consider N-1 and 0 as consecutive integers because according to the distance measure specified by (2), the distance between them is 1.
 It is clear that the minimum spread of an interleaver is 2.

 Clockwise   and counterclockwise $w$-sequences have been defined  in \cite{Abra}. Clockwise $w$-sequences mean sequences
such as $\{0,1,..w-1\},\{1,2,..w\},\{N-2,N-1,0,1,...w-3\}$ etc. while counterclockwise $w$-sequences mean sequences
such as $\{w-1,w-2,..0\},\{w,w-1,...1\},\{w-3,w-2,...0,N-1,N-2\}$ etc.  The event of
a permutation having spread more than 2 can be identified with the event of a permutation without any clockwise or counterclockwise 2-sequence. The number of permutations
without any clockwise or counterclockwise $w$-sequences, $M_{0}(N,w)$ is also given in 
\cite{Abra}. We can find out the number of permutations without any clockwise or counterclockwise 2-sequences from that:
\begin{eqnarray}
M_{0}(N,2) 
&=& N!+ \sum_{i=1}^{N-1} (-1)^{i}\sum_{a=1}^{i} 2^{a}\frac{N}{N-i}\nonumber\\
&&.{{i-1}\choose{a-1}}{{N-i}\choose{a}} (N-i)!
\end{eqnarray}
Thus $M_{0}(N,2)$ is the number of interleavers of blocklength $N$ with spread more than 2. A total of $N!$ interleavers are there. Say, $K_{s}(N)$ is the number of interleavers of blocklength $N$ with spread $s$. So the number of interleaver with spread equal to 2 is given by,
\begin{equation}
 K_{2}(N)= N!-M_{0}(N,2)
\end{equation}
From (3), the probability that a randomly chosen interleaver
 will have  spread more than 2 is given by,
\begin{eqnarray}
P(spread >2)&=&\frac{M_{0}(N,2)}{N!}\nonumber\\
&=& 1+ \sum_{i=1}^{N-1} (-1)^{i}\sum_{a=1}^{i} 2^{a}\frac{N}{N-i}\nonumber\\
&&.{{i-1}\choose{a-1}}{{N-i}\choose{a}}. \frac{(N-i)!}{N!}
\end{eqnarray}
Now consider the case when  $N\rightarrow \infty$ :
\begin{eqnarray}
P(spread >2)&=& 1+ \lim_{N\rightarrow \infty}\sum_{i=1}^{N-1} (-1)^{i}\sum_{a=1}^{i} 2^{a}\frac{N}{N-i}\nonumber\\
&&.{{i-1}\choose{a-1}}{{N-i}\choose{a}}. \frac{(N-i)!}{N!}\nonumber\\
&=& 1+ \sum_{i=1}^{\infty}(-1)^{i}\sum_{a=1}^{i} 2^{a} {{i-1}\choose{a-1}}\nonumber\\
&& .\lim_{N\rightarrow \infty} [\frac{N}{N-i} {{N-i}\choose{a}} \frac{(N-i)!}{N!}]  \nonumber\\
&=&1+ (-1)^{i}\sum_{a=1}^{i}\frac{2^{a}}{a!} {{i-1}\choose{a-1}}\nonumber\\
&&.\lim_{N\rightarrow \infty}\left [\frac{N}{N-i}.\frac{(N-i)!}{(N-i-a)!}\right .\nonumber\\
&& .\left .\frac{(N-i)!}{N!}\right ]
\end{eqnarray}



For $a< i$ the degree of N in each term inside the limit in (6) is negative and these terms
go  to zero as $N\rightarrow \infty$. The term with $a=i$ goes to 1 as $N\rightarrow \infty$. 

So, from (6) as $N\rightarrow \infty$,
\begin{equation}
P(spread >2)= 1+ \sum_{i=1}^{\infty}\frac{(-2)^{i}}{i!} = e^{-2}
\end{equation}

So, as  $ N\rightarrow \infty, \quad P(spread=2)= 1-e^{-2}=0.8647$.

\quad

Thus it is fair to state that if we randomly pick an interleaver it is very likely that it
will have  spread 2. At the same time finding an interleaver with spread more than 2 is not very hard.
In every $e^{2}\approx 8$ random interleavers, there is likely to be one with spread more than 2. Note
that here we are considering  large blocklengths. For finite blocklengths the probability that an interleaver
will have  spread more than 2 is lesser. 

\begin{figure}
\centering
\includegraphics[width=3.5in]{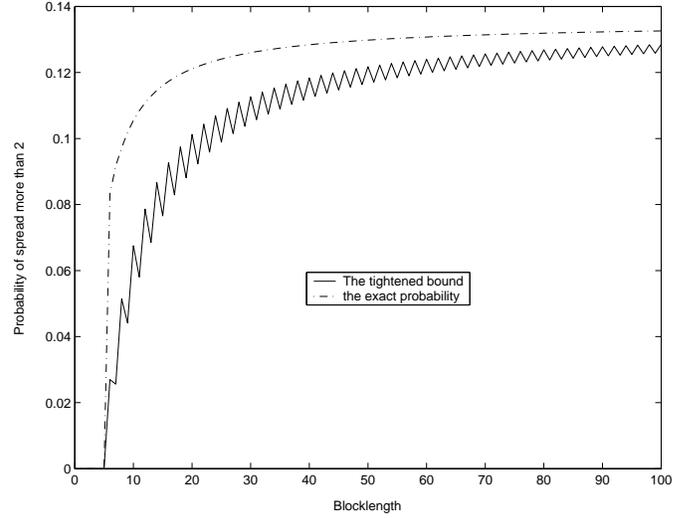}
\caption{The upper curve shows the exact probability of finding an interleaver with spread more than 2 as a function of blocklength. The lower curve gives the lower bound on the same, obtained from the tightened bound.}
\label{label2}
\end{figure}

The  plot of  (6) is shown in Fig.\ \ref{label2} (the upper curve). It  shows that the probability, that an interleaver
will have  spread more than 2 gets very close to $e^{-2}$ for blocklengths as small as 100.

\quad

\section{Interleavers With Higher Spread}

\quad

{\bf Theorem:} If we randomly choose an interleaver from the set of all possible interleavers of blocklenth $N$ then,
\begin{equation}
P(spread\geq s)\geq \left[\frac{(N-2s+3)^{s-1}.(N-2s+2)^{s-1}}{(N-1)(N-2)...(N-2(s-1))}\right]^{N}
\end{equation}

{\bf Proof:}

Let $A_{s}$ correspond to the event that a randomly picked interleaver has  spread at least $s$.

Let $A_{is}$ be the event that for any fixed $i$
\begin{eqnarray}
|i-j|_{N}+|\pi(i)-\pi(j)|_{N}&\geq & s \nonumber\\
\forall \,\,  0\leq j\leq N-1,&& j\neq i
\end{eqnarray}
It is clear that
 \begin{equation}
 P(A_{s})= P(\bigcap_{i=0}^{N-1} A_{is})
 \end{equation}

It was proved in \cite{Sphere} that for finite blocklength $N$ spread is upper bounded by $\lfloor \sqrt{2N}\rfloor$.
 Let us denote the maximum value of spread for blocklength $N$ as $s_{max}$. Now, $\bigcap_{i=0}^{N-1} A_{is} \neq \phi$ only if $s\leq s_{max}$,
 otherwise $P(A_{s})=0$.

Let us take any $i$. $\pi(i)$ will be equal to $k$ with probability $\frac{1}{N}$. Let us take the case
where $\pi(i)=k$. Now the interleaver $\pi$ will have  spread more than $s$ if

\begin{eqnarray}
\left|i-j\right|_{N}+\left|k-\pi (j)\right|_{N} & \geq &  s,\nonumber\\
\forall \,\,  0\leq j\leq N-1, && j\neq i
\end{eqnarray}

Let us call the above event $A_{iks}$. Clearly
\begin{eqnarray}
P(A_{is})&=& P(A_{i0s})P(\pi (i)=0) + P(A_{i1s})P(\pi (i)=1)\nonumber\\
&& +P(A_{i2s})P(\pi (i)=2)+ ...\nonumber\\
&& .....+P(A_{i(N-1)s})P(\pi (i)=N-1)\nonumber\\
&=& \frac{1}{N}\sum_{k=0}^{N-1}P(A_{iks})
\end{eqnarray}

\begin{figure}
\centering
\includegraphics[width=3.5in]{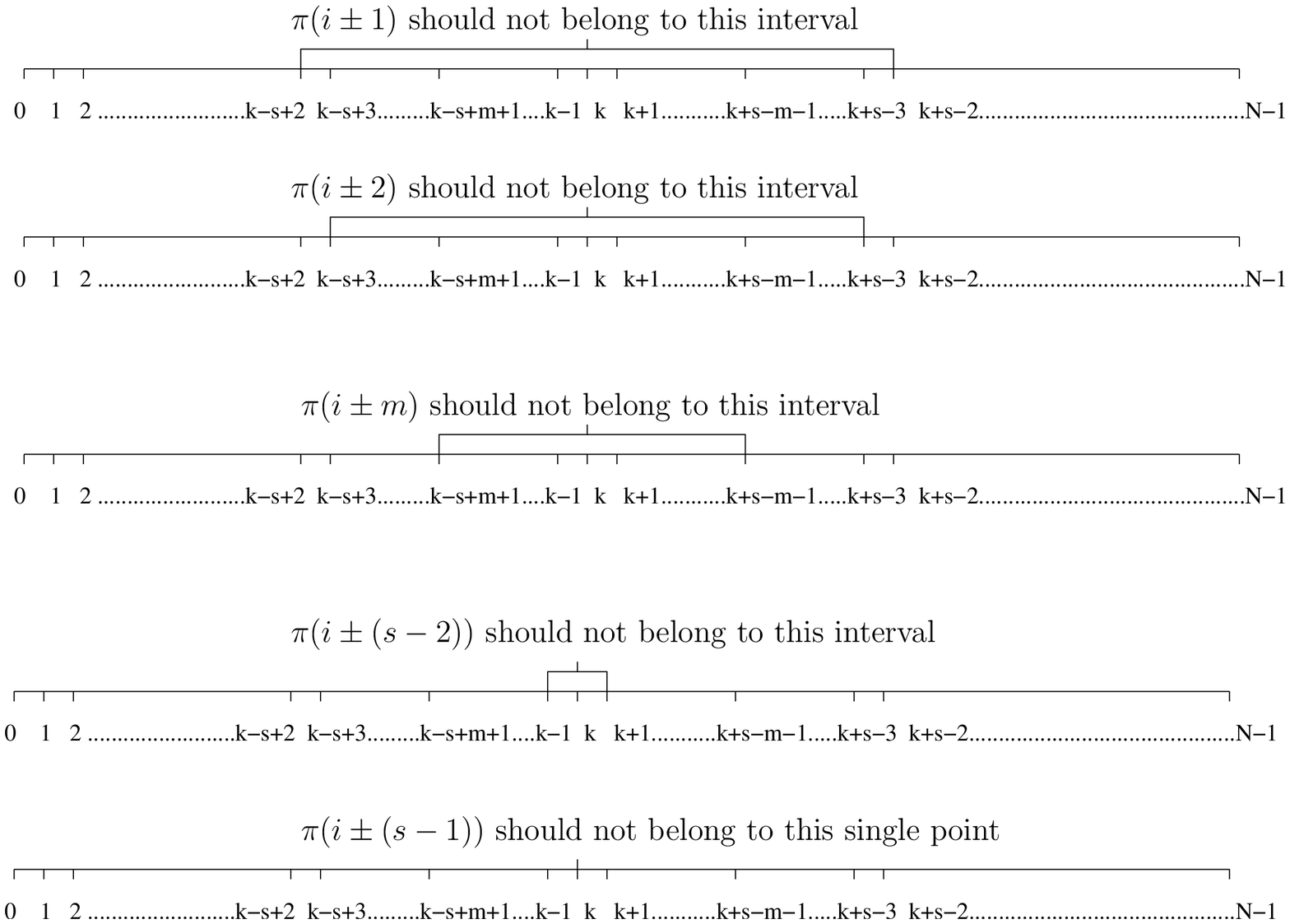}
\caption{This shows the conditions for which $A_{iks}$ will be true. Here $\pi(i)=k$. So the conditions should ensure $\left|i-j\right|_{N}+\left|k-\pi (j)\right|_{N}\  \geq \  s$ }
\label{label_name}
\end{figure}

As mentioned earlier $0$ and $N-1$ are considered  as neighbors and hence
 the entire arrangement of {0,...N-1} can be thought of as having a circular structure. So whatever may be the
value of $i$ or $k$ (even if they are near $0$ or $N-1$) the event $A_{iks}$ will have the same
probability. For the purpose of convenience in  illustration   we have taken $i$ and $k$ sufficiently away
from $0$ or $N-1$ in Fig.\ \ref{label_name}.

Fig.\ \ref{label_name}  shows the conditions for which the event $A_{iks}$ will be true. If $ \pi(i)=k$ then to have spread at least $s$, $\pi(i\pm 1)$ should not belong to the integer interval  from $k-s+2$ to $k+s-2$, $\pi(i\pm 2)$ should not belong to the integer interval  from $k-s+3$ to $k+s-3$, and so on, as shown in Fig. \ref{label_name}.

Say an integer interval  from $p$ to $q$ is written as $\{p,q\}$. Then, as explained in  Fig.\ \ref{label_name}, the event $A_{iks}$ will be true if
\begin{eqnarray}
\pi(i\pm m)\notin \{k-s+m+1,k+s-m-1\}\nonumber\\
\quad\forall \quad1\leq m\leq s-1
\end{eqnarray}

For a fixed $m$, let us call the event of (13) as $A_{iksm}$. It is clear that,
\begin{equation}
P(A_{iks})= P(\bigcap_{m=1}^{s-1} A_{iksm}).
\end{equation}

Now,
\begin{eqnarray}
&&P(A_{iksm}) \nonumber\\
&=& P([\pi(i+m)\notin \{k-s+m+1,k+s-m-1\}]\nonumber\\
&&\bigcap[\pi(i-m)\notin \{k-s+m+1,k+s-m-1\}])  \nonumber\\
&=& P(\pi(i+m)\notin \{k-s+m+1,k+s-m-1\}) \nonumber\\
& &.P([\pi(i-m)\notin \{k-s+m+1,k+s-m-1\}]\nonumber\\
& &|[\pi(i+m)\notin \{k-s+m+1,k+s-m-1\}])  \nonumber\\
&=&\frac{N+1-2(s-m)}{N-1}.\frac{N+1-2(s-m)-1}{N-2}
\end{eqnarray}

In a similar way, we can find out the conditional probability  $P(A_{iksm}|A_{iks1}.A_{iks2}....A_{iks(m-1)})$. It will be similar to (15), but as $2(m-1)$ points are already occupied because of $A_{iks1},A_{iks2}....A_{iks(m-1)}$ given, $2(m-1)$ should be subtracted from the numerators and denominators of both fractions of (15), i.e.,
\begin{eqnarray}
&&P(A_{iksm}|A_{iks1}.A_{iks2}....A_{iks(m-1)}) \nonumber\\
& =&\frac{N+1-2(s-m)-2(m-1)}{N-(2m-1)}\nonumber\\
&&.\frac{N+1-2(s-m)-2(m-1)-1}{N-(2m-1)-1} \nonumber\\
&= &\frac{(N-2s+3)(N-2s+2)}{(N-2m+1)(N-2m)}.
\end{eqnarray}

From (14),
\begin{equation}
P(A_{iks})= \prod_{m=1}^{s-1} P(A_{iksm}|A_{iks1}.A_{iks2}....A_{iks(m-1)})
\end{equation}
\begin{eqnarray}
& =& \prod_{m=1}^{s-1} \frac{(N-2s+3)(N-2s+2)}{(N-2m+1)(N-2m)}  \quad \mbox{from (16)} \nonumber\\
& =& \frac{(N-2s+3)^{s-1}.(N-2s+2)^{s-1}}{(N-1)(N-2)...(N-2(s-1))}.
\end{eqnarray}

\quad

We see that the right hand sides of (14), (16) and (18) are all independent of $k$, because whatever may be the value of $\pi(i)$, the probabilities are same. Now, from (12),
\begin{eqnarray}
&&P(A_{is})\nonumber\\
&=& \frac{1}{N} \sum_{k=0}^{N-1}\frac{(N-2s+3)^{s-1}.(N-2s+2)^{s-1}}{(N-1).(N-2)...(N-2(s-1))}\nonumber\\
&=&\frac{(N-2s+3)^{s-1}.(N-2s+2)^{s-1}}{(N-1)(N-2)...(N-2(s-1))}.
\end{eqnarray}

So, as expected $P(A_{is})$  is same for all $i$'s.

 We can write (10) as,
\begin{equation}
P(A_{s})=\prod_{i=0}^{N-1}P(A_{is}|A_{0s}.A_{1s}....A_{(i-1)s})
\end{equation}

If $s> s_{max}$ then at least for some $i,\quad P(A_{is}|A_{0s}.A_{1s}....A_{(i-1)s})=0$.
But if interleaver with  spread $s$ exists, then  it can be argued that the conditional probability of $A_{ls}$, if  $A_{is}$ is true for $i=0,1,..l-1$, is larger than or equal to the unconditional probability of $A_{ls}$. Hence,
\begin{equation}
P(A_{is}|A_{0s}.A_{1s}....A_{(i-1)s}) \geq P(A_{is}) \quad \mbox{for all}\quad 0\leq i\leq N-1
\end{equation}

Thus we get the following inequality from (20) and (21), for $s\leq s_{max}$ :
\begin{eqnarray}
&&[P(A_{is})]^{N} \leq P(A_{s}) \\
&\mbox{or}&\left [\frac{(N-2s+3)^{s-1}.(N-2s+2)^{s-1}}{(N-1).(N-2)...(N-2(s-1))}\right ]^{N} \leq P(A_{s})\nonumber\\
\quad
\end{eqnarray}

This completes the proof of the theorem.

\quad

\section{Tightening the Bound}

\quad

 If the event $A_{is}$ is true for $p-(s-1)< i <p$ and $p+(s-1)>i>p$, then $A_{is}$ will be true for $i=p$ as well. So to have the event $A_{s}$, we need  $A_{is}$ to be satisfied for all $i$'s except every  $(s-1)^{th}$, because they will be then be automatically satisfied. Using this observation we now tighten the bound given in (8). 
Thus (20) can be written as,
\begin{eqnarray}
&&P(A_{s})=\prod_{i=0}^{s-2}P(A_{is}|A_{0s}.A_{1s}....A_{(i-1)s})\nonumber\\
&&.\prod_{i=s}^{2s-2}P(A_{is}|A_{0s}.A_{1s}....A_{(s-2)s}A_{(s)s}....A_{(i-1)s})\nonumber\\
&&.\prod_{i=2s}^{3s-2}P(A_{is}|A_{0s}...A_{(s-2)s}A_{(s)s}..A_{(2s-2)s}A_{(2s)s}..A_{(i-1)s})\nonumber\\
&&..\prod_{i=ks}^{N-1}P(A_{is}|A_{0s}..A_{(s-2)s}A_{(s)s}..A_{(2s-2)s}A_{(2s)s}...\nonumber\\
&&...A_{(i-1)s})\nonumber\\
&& .P(A_{(s-1).s}|A_{0s}...A_{(s-2)s}A_{(s)s}...A_{(2s-2)s}A_{(2s)s}..\nonumber\\
&&..A_{(N-1)s}) \nonumber\\
&&.P(A_{(2s-1).s}|A_{0s}.A_{1s}.....A_{(2s-2)s}A_{(2s)s}....A_{(N-1)s})\nonumber\\
&&....P(A_{(ks-1).s}|A_{0s}.A_{1s}.......A_{(N-1)s})
\end{eqnarray}

\quad

where $k$ is an integer such that $ks\leq N-1$ but $(k+1)s >N-1$.
 In the last $\lfloor{\frac{N}{s-1}}\rfloor$ terms of (24) the probability of each $(s-1)^{th}$ element satisfying $A_{is}$ is expressed given all 
other elements satisfying  $A_{is}$. Clearly  each of these will be equal to unity. So (24) is simplified to
\begin{eqnarray}
&&P(A_{s})=\prod_{i=0}^{s-2}P(A_{is}|A_{0s}.A_{1s}....A_{(i-1)s})\nonumber\\
&&.\prod_{i=s}^{2s-2}P(A_{is}|A_{0s}.A_{1s}....A_{(s-2)s}A_{(s)s}....A_{(i-1)s})\nonumber\\
&&.\prod_{i=2s}^{3s-2}P(A_{is}|A_{0s}...A_{(s-2)s}A_{(s)s}..A_{(2s-2)s}A_{(2s)s}.\nonumber\\
&&...A_{(i-1)s})\nonumber\\
&&..\prod_{i=ks}^{N-1}P(A_{is}|A_{0s}..A_{(s-2)s}A_{(s)s}..A_{(2s-2)s}A_{(2s)s}..\nonumber\\
&&....A_{(i-1)s})\nonumber\\
\quad
\end{eqnarray}

  Using an  argument similar to (21),  and the fact that there are $N-\lfloor{\frac{N}{s-1}}\rfloor$ terms to be multiplied in (25), we get for $s\leq s_{max}$:

\begin{eqnarray}
&&[P(A_{is})]^{N-\lfloor{\frac{N}{s-1}}\rfloor} \leq P(A_{s})\nonumber\\
&\mbox{or}&\left [\frac{(N-2s+3)^{s-1}.(N-2s+2)^{s-1}}{(N-1)(N-2)...(N-2(s-1))}\right ]^{N-\lfloor{\frac{N}{s-1}}\rfloor}\nonumber\\
&& \leq P(A_{s})
\end{eqnarray}
which is tighter than (8).

\quad

\begin{figure}
\centering
\includegraphics[width=3.5in]{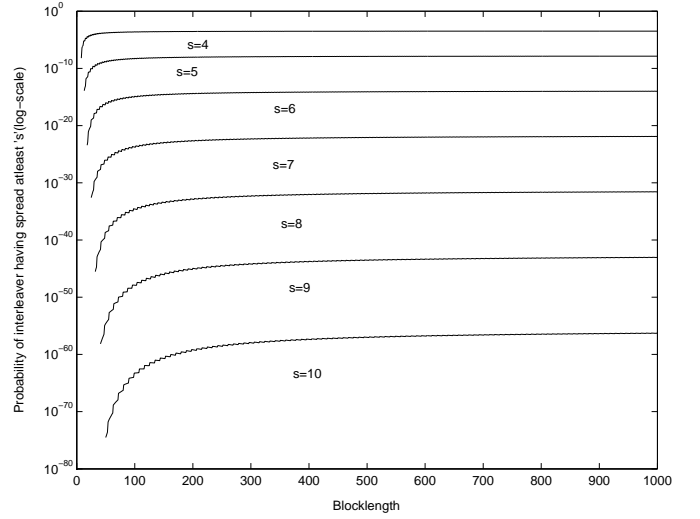}
\caption{Lower bounds on probability of spread at least s vs blocklength from (26), these bounds reach constant values as blocklength increases}
\label{fig_sim}
\end{figure}

If we multiply the left hand side of (26) with $N!$ we will get a lower bound on the number of interleavers with spread at least $s$.  
\begin{eqnarray}
&&\sum_{p\geq s} K_{p}(N)=N!P(A_{s})\nonumber\\
&&\geq N!\left [\frac{(N-2s+3)^{s-1}.(N-2s+2)^{s-1}}{(N-1)(N-2)...(N-2(s-1))}\right ]^{N-\lfloor{\frac{N}{s-1}}\rfloor}\nonumber\\
\quad
\end{eqnarray}

Expanding (26) and writing it as a product of $2(s-1)$ fractions  we get,
\begin{eqnarray}
P(A_{s}) &\geq &  \left [\frac{N-2s+3}{N-1}.\frac{N-2s+3}{N-2}....\frac{N-2s+3}{N-(s-1)}\right .\nonumber\\
&& \left . \frac{N-2s+2}{N-s}....\frac{N-2s+2}{N-2(s-1)}\right ]^{N-\lfloor{\frac{N}{s-1}}\rfloor}\nonumber\\
&=&(1-\frac{2s-4}{N-1})^{N-\lfloor{\frac{N}{s-1}}\rfloor}...(1-\frac{s-2}{N-s+1})^{N-\lfloor{\frac{N}{s-1}}\rfloor}\nonumber\\
&& (1-\frac{s-2}{N-s})^{N-\lfloor{\frac{N}{s-1}}\rfloor}...1   
\end{eqnarray}
To check what the bound turns out for large $N$, we take $N\rightarrow\infty$. Using the result 
\begin{equation}
\lim_{x\rightarrow \infty}(1-\frac{a}{x})^{bx}= e^{-ab}
\end{equation}
 we get
\begin{eqnarray}
P(A_{s})&\geq& e^{-(2s-4)\frac{s-2}{s-1}}....e^{-(s-2)\frac{s-2}{s-1}}.e^{-(s-2)\frac{s-2}{s-1}}....1 \nonumber\\
 &=& e^{-(s-2)\frac{s-2}{s-1}}.[e^{-[0+1+2+...+(2s-4)]\frac{s-2}{s-1}}] \nonumber\\
&=& e^{-2(s-2)^{2}}
\end{eqnarray}


\quad

Thus the probability of a randomly chosen interleaver having  spread more than 2 (at least 3) for large blocklengths is lower bounded by $e^{-2(3-2)}=e^{-2}$, from (30). This bound exactly matches the result  in (7). The bound in (26) is plotted in Fig.\ \ref{label2}  (the lower curve) for $s=3$. We can see from the figure that the tightness of the bound in (26) for $s=3$, improves as blocklength increases. In the plot of the bound in Fig. \ref{label2}, the ripples come because of the floor function in the expression of (26). It can be noticed that these ripples gradually decrease as the blocklength increases.

\begin{figure}
\centering
\includegraphics[width=3.5in]{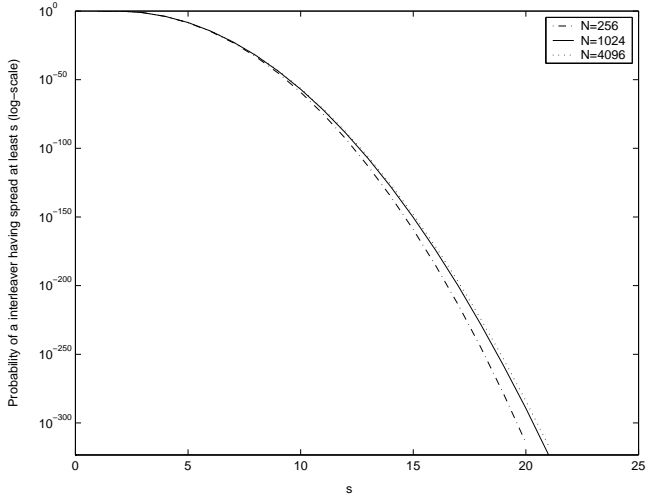}
\caption{Lower bounds on probability of spread at least s is shown as a function of s (from (26)) for different blocklength N.}
\label{label3}
\end{figure}

We have plotted the probability bounds of (26) as a function of $N$ in Fig.\ \ref{fig_sim}. It can be seen that the ripples which were prominent in the plot of the bound for $s=3$ in Fig.\ \ref{label2}, are present in these plots also, but they are barely visible. We observe that each curve initially rises but then very quickly becomes almost horizontal indicating convergence to a limiting value. Further, as expected the curves for higher values of $s$ are much below the curves for smaller value of $s$. The extremely small values of the probabilities need not necessarily mean that the number of interleavers with high spread is negligibly small because a lower bound on the number of interleavers is obtained by  multiplying  these probabilities by $N!$.

 The lower bounds of (26) are plotted in Fig.\ \ref{label3} as  functions of $s$  for blocklengths 256, 1024 and 4096. We can see that the bounds for different blocklengths are quite close to each other. There is very little separation  between  the curves for $N=1024$ and $N=4096$. So for large blocklengths these bounds are almost independent of blocklength.  From the values in the plots it is clear that if we randomly pick up interleavers, it is  difficult to find  an
interleaver with high spread  as the size of the search space needed to guarantee this  is very large (search space size has the order $e^{2(s-2)^{2}}$). For blocklengths larger than 1000, approximately $86.47\%$ interleavers have  spread 2. At least $13.53\%$ interleavers have  spread more than 2, at least $0.0335\%$ interleavers have  spread more than 3 and at least $1.52 \mbox{x} 10^{-6}\%$ have  spread more than 4. For large blocklengths, the fraction  of interleavers with spread more than $s$ decays as fast as $e^{-2(s-2)^{2}}$. Thus, it is fair to state that the expected spread of a  random interleaver is a little more than 2.  

\quad

\section{Conclusion}

\quad

We have addressed the problem of determining the number of interleavers of a given blocklength $N$ that will have  spread $s$. It has been possible to determine the exact expression for this number when $s$ is equal to two. For other values of $s$ lower bounds have been obtained on the number of interleavers having  spread at least  $s$. 
  The probability that a randomly chosen interleaver will have spread two, converges quickly to $1-e^{-2}$, as blocklength increases. The lower bound on the probability  that a random interleaver will have spread at least s, converges to  $e^{-2(s-2)^{2}}$. 

\quad

\end{document}